\documentclass[10pt,journal,compsoc]{IEEEtran}

\usepackage[utf8]{inputenc}
\usepackage{geometry}
\usepackage{url}
\usepackage{graphicx}

\usepackage{listings}
\usepackage{fancyvrb}

\usepackage{tgcursor}

\usepackage{cite}

\newcommand{\Gaudi }{\textsc{Gaudi }}
\newcommand{\DaVinci }{\textsc{DaVinci }}

\begin{document}
\hyphenation{ana-lysis ana-lyses}

\title{Provenance tracking in the LHCb software}

\author{Ana~Trisovic,
        Chris~R.~Jones,
        Ben~Couturier,
        and~Marco~Clemencic%
\IEEEcompsocitemizethanks{
\IEEEcompsocthanksitem A. Trisovic was with the University of Chicago, 1100 E 57th St, Chicago, IL 60637 and CERN - European Organization for Nuclear Research 1, Esplanade des Particules, CH-1217 Meyrin, Switzerland.\protect\\
Email: anatrisovic@g.harvard.edu
\IEEEcompsocthanksitem C.~R.~Jones is with the Cavendish Laboratory, University of Cambridge, JJ Thomson Ave, Cambridge CB3 0HE, United Kingdom.
\IEEEcompsocthanksitem B.~Couturier~and~M.~Clemencic are with CERN - European Organization for Nuclear Research 1, Esplanade des Particules, CH-1217 Meyrin, Switzerland.}%
}

\IEEEtitleabstractindextext{%
\begin{abstract}
Even though computational reproducibility is widely accepted as necessary for research validation and reuse, it is often not considered during the research process. This is because reproducibility tools are typically stand-alone and require additional training to be employed. In this paper, we present a solution to foster reproducibility, which is integrated within existing scientific software that is actively used in the LHCb collaboration. Our provenance tracking service captures metadata of a dataset, which is then saved inside the output data file on the disk. The captured information allows a complete understanding of how the file was produced and enables a user to reproduce the dataset, even when the original input code (that was used to initially produce the dataset) is altered or lost. This paper describes the implementation of the service and gives examples of its application.
\end{abstract}

}

\maketitle

\IEEEraisesectionheading{\section{Introduction}\label{sec:introduction}}

\IEEEPARstart{R}{eproducibility} of research is accepted as necessary for result validation, sharing and reuse~\cite{chen2018open, national2019reproducibility}. Dozens of standalone tools have been created to preserve research materials and thus enable reproducibility. Some of these tools help with documentation~\cite{chen2018open}, the others with tracking system dependencies thus allowing cross-platform compatibility~\cite{boettiger2015introduction}, some even capture the whole analysis in an automatized workflow~\cite{goecks2010galaxy}. A number of reproducibility tools are web-based with a user-friendly graphical interface~\cite{staubitz2016codeocean}. When employed in the research process, they are indeed able to solve the problems of reproducibility and allow reuse. However, reproducibility-focused research is still not common in scientific communities, and a small number of people who use the tools represent an exception. There are several reasons why these tools, although useful, are not ubiquitously used. First, researchers are often working under constant pressure to present results and publish papers~\cite{baker2016there}, which can cause lack of time to employ reproducibility practices in their everyday work. Often in research groups, many computationally intensive tasks are done by students or young scientists with typically fixed-term stay, during which they might not manage to ensure reproducibility of their work. Second, there is often a lack of training on best practices in data analysis in social and (some) natural sciences, even though these disciplines also face an expansion of data and the need for computationally intensive analyses. Researchers in those fields are hence hesitant to use more software than they need to, including the reproducibility tools. Because of this, the tools are not normally used in everyday scientific work, which can thus hinder research reproducibility and reuse at a later stage.

In this paper, we argue that the best way to foster reproducibility is to integrate it within existing scientific software that is already in use. This way enabling reproducibility is simple and it seamlessly integrates into daily researchers' work. It lessens the burden of the analysts who might otherwise need to install and learn to use a new software tool. We demonstrate the proposed solution through our work at LHCb, the high-energy physics experiment at CERN. According to a survey conducted at the experiment~\cite{trisovic2018data}, most LHCb analysts use the official software to fully or in part perform their physics analysis. These results served as a starting point for addressing the reproducibility challenge. Our solution is designed and developed within the official experimental software to capture data provenance, which is then saved inside the output data file on the disk. The stored provenance allows understanding how a file was produced~\cite{pasquier2017if} and provides sufficient information to entirely reproduce the dataset, eliminating the need for the original input code or even documentation. 

\section{The LHCb software}

The LHCb software is the essential component in all data-related activities at LHCb. It is used in a range of data processing environments, from real-time data collection in the experimental setup, data reconstruction, simulation, to advanced physics analyses and collision visualization. The LHCb software is based on the C++ object-oriented framework called \textsc{Gaudi}~\cite{barrand2001gaudi}, which provides a common infrastructure and environment for the software applications of the experiment. \Gaudi was used during both the first and second run of the LHC and it proved to be a reliable and flexible platform for LHCb data processing on the CERN Computing Grid. The project was created in 1998 by the LHCb collaboration, but since then it was released open-source~\footnote{\url{https://gitlab.cern.ch/lhcb/Gaudi}} and adopted by a number of other high-energy physics experiments such as ATLAS~\cite{1742-6596-219-4-042006}.

The software is organized in a modular architecture of smaller and more manageable packages. Each module has a defined functionality and an interface through which it interacts with the other components. Such architecture provides layers of abstraction for developers, meaning that one does not need to understand the whole framework to contribute to one of the components.

\begin{figure*}[ht]
\centering
   \includegraphics[width=\textwidth]{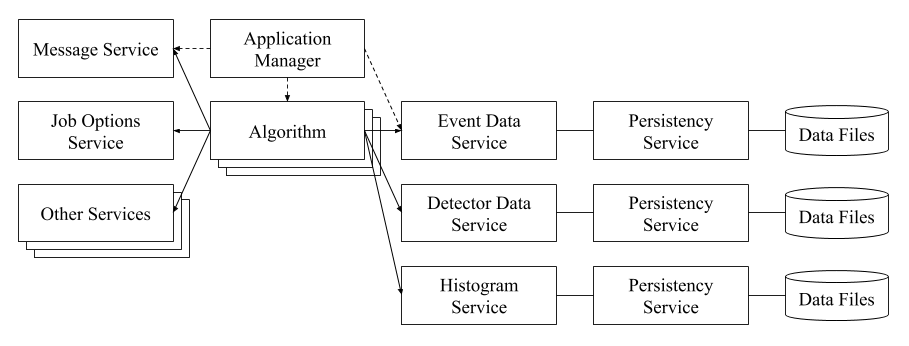}
   \caption{A snapshot of \Gaudi during runtime shows the basic modules and the connections between them.} \label{fig:gaudia}
\end{figure*}

The basic modules of the \Gaudi framework and the connections between them are shown in Figure \ref{fig:gaudia}. The Application manager (in code denoted as \texttt{ApplicationMgr}) controls the execution of the jobs within the framework. It creates and initializes the required modules in the system, and retrieves input data. The input data is a collection of highly-structured information that describe particle collisions (also called events) recorded inside the detectors or created in simulations. The Application manager loops over the input data events and executes the algorithms. If there are any errors in the execution, the Application manager will handle them and in the end it will terminate the job. The module Algorithms (\texttt{Algorithms}), has a central place in the job execution, as it performs data processing on all input data. At the end, output files are produced in a form of another data file or other type of output. 

The \Gaudi services provide various utilities and services for the Algorithms in the system, which are also initialized by the Application manager at the beginning of a job. Normally, only one instance of a service is required in the job. There are a number of different services within the framework that can be used by the Algorithms but some of the main ones are:

\begin{itemize}
    \item Event Data service (\texttt{EventDataSvc}) and Histogram service \\(\texttt{HistogramDataSvc})  that read and process individual collision events,
    \item Detector Data service (\texttt{DetDataSvc}) for capturing detector data,
    \item Message service (\texttt{MessageSvc}) logs progress or errors in the Algorithms and
    \item Tool Service (\texttt{ToolSvc}) manages Algorithm tools, which are required during the Algorithm execution. 
\end{itemize}

The Persistency services allow writing the output data on the disk as presented in the figure. Finally, there are many other services in the framework that provide specialized functions that can be enabled and disabled by the users. Each of the service classes is used by the Algorithms via an interface, which is a helper class that defines the functionality of a service through a number of public methods. These methods also allow the communication of the service with other components of the framework.

In order to initiate these services, \Gaudi ingests a set of configurations defined by physics analysts in (one or more) Python files at the beginning of a data processing job. The Python files are called \textit{Python application configuration files} and they specify how the job will run and designate configurations to required C++ objects in \textsc{Gaudi}. These configurations are passed by the Job Options Service (\texttt{JobOptionsSvc}) and applied at runtime. Every Algorithm used in the framework is configured using these options.

\section{Implementation of the \\Provenance tracking service}

The provenance tracking service, called Metadata service (\texttt{MetaDataSvc}), is implemented as a \Gaudi service inside the module \Gaudi services (\texttt{GaudiSvc}). Its functionality is simple: collect information about a job and capture it in an object, which is then stored as metadata in the output data file. There is a number of LHCb data formats in which an output of a \Gaudi job can be stored. Data formats that are most commonly used in high-energy physics analysis are based on the ROOT file format~\cite{brun1997root}. A ROOT file format is very flexible and it acts like a UNIX file directory, meaning that it can store directories and data objects organized in an arbitrary order. Furthermore, it can store any C++ object, like for example histograms, plots, tables and other. The metadata object, named {\it info}, is implemented as a dictionary (a map \texttt{std::map} in C++), which is a general-purpose data structure for storing a group of objects. A dictionary captures a set of key-value pairs, which, in this implementation, are names of application properties as keys, and their configurations as values.

The service is implemented in a C++ class with the following main methods that execute its workflow in runtime:
 
\begin{enumerate}
    \item \texttt{isEnabled} captures information whether the service is enabled in the job or not.
    \item \texttt{start} initiates the service and calls \texttt{collectData}.
    \item \texttt{collectData} executes the main functionality of the service. It traverses and queries the \Gaudi tools (\texttt{ToolSvc}), services (\texttt{Services}), algorithms (\texttt{Algorithms}) and Job Options (\texttt{JobOptionsSvc}) to capture their configurations. \item \texttt{getMetaData} returns the object that stores a dictionary of the job configurations.
\end{enumerate}

Within the \Gaudi framework, there are three audit methods that follow a job execution. They are automatically invoked by the Application manager at the start of every job. Those methods are:

\begin{itemize}
    \item \texttt{initialize} that initializes algorithms and services, and applies job options,
    \item \texttt{execute} that executes the main function of the job,
    \item \texttt{finalize} that is called at the end of the job.
\end{itemize}

The Metadata service is called and initialized from the \texttt{finalize} method, at the point when all configurations have been correctly applied to the job. During the execution of the \texttt{initialize} or \texttt{execute} methods only a selection of the job configurations (not all) have been applied, meaning that calling the service from these methods would cause an information loss. Therefore, the metadata is only captured at the end of the job execution at the moment when the output ROOT file is written to the disk. The service functionality is finished when the metadata dictionary is saved into the file.   
 
\section{Using the Provenance \\tracking service}

Even though LHCb data analyses can be done using a wide range of tools and programming languages, retrieving the data from the CERN Computing Grid needs to be done using the LHCb software. Typically an LHCb application called \textsc{DaVinci} is used in this step. \DaVinci is a physics analysis application based on the \Gaudi framework that is primarily used as a part of data processing to calculate a variety of kinematic quantities for recorded particles, but it is also used in fine data selections to extract particle decays of interest. As any other \textsc{Gaudi}-based physics application that is used in LHCb, \DaVinci can access and utilize the Metadata service available in the framework. 

Capturing data provenance within the \Gaudi framework is the simplest way for the analysts to conduct reproducible analysis. The provenance captured with the Metadata service is useful in a number of different scenarios: 

\begin{enumerate}
    \item The first scenario refers to reproducing a dataset using the original application version. A common practice in using the LHCb software is to use the latest available version at the time, which is done by specifying the keyword 'latest' in the code. The latest version is recommended for use because it captures recent developments with new features or solutions to known bugs. However, by doing this the analysts do not necessarily record what application version they used, which may at the later stage, when the ``latest'' version changes, hinder reproducibility. Even though the applications run on the same framework, two different versions can produce different datasets. For example, the application \textsc{DaVinci v39r0} and \textsc{DaVinci v42r3} may produce slightly different outputs even when they ingest the same application configuration file. The Metadata service hence facilitates reproducibility by storing the exact information about the version of the software used, which users can unequivocally specify when reproducing the dataset.
    \item Another common practice in conducting physics analyses is to use one application configuration file to produce many similar datasets with slight configuration changes. This is typical when for example an analyst wants to test different selections on the data. Such practice produces a number of data files without clear information how they were produced, and if an analyst by mistake mislabels them (by marking, for example, the origin year of data 2012 instead of 2011) mistakes in the physics analysis may happen. The Metadata service facilitates reproducibility in this scenario as it provides information (that is not prone to human error) about these slight configuration changes in the dataset and thus it enables the analyst to double-check their work.
    \item Furthermore, working with collaborators on an analysis typically means sharing disk workspace on a CERN or University server. Each of the collaborators creates temporary or derived data files, and often after some time they forget how and why these files were created. This hinders reproducibility as it conceals potential steps that were previously taken in the analysis workflow. The Metadata service captures what were the inputs and processes conducted on each dataset, which could help the analysts to reconstruct the analysis workflow and distinguish between temporary (testing) and requisite (final) datasets.
    \item Finally, if a bug in one application version is identified, the analysts need to know whether they had created datasets using this version as it could negatively affect their analysis results. If the provenance of these datasets is available, they could instantly evaluate whether they used the faulty version and recreate the datasets with another application version. In this scenario, the information provided by the Metadata service helps in troubleshooting, result verification and ensuring trustworthiness.
\end{enumerate}

The Metadata dictionary provides clear information what application version was used and how it was used, thus avoiding the ambiguity illustrated in these scenarios. Our solution can be immensely useful for everyday development and validation of research.

In the current implementation, the service needs to be manually enabled in order to be used in data processing. This is done by including just one additional line of code in the Python application configuration file, that appends the Metadata service into the list of existing services handled by the Application manager. Once the Python application configuration file is completed, it is passed to \Gaudi to run the job. 

An output dataset in the ROOT file format that captures data of a particle decay $D^+\rightarrow K^+\mu+\mu-$ (D2Kmumu) is shown in Figure~2. The dataset was created using the application \textsc{DaVinci v42r3}. This information and other configurations are captured in the additional {\it info} object, which is also saved and visible in the figure. 

\begin{figure}[ht]
    \centering
    \includegraphics[width=.5\textwidth]{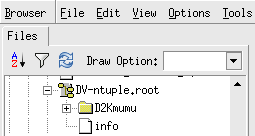}\label{fig:mdsvc1}
    \caption{After data file \textsc{DV-ntuple.root} has been produced with the provenance tracking service, it stores an additional object with the job metadata called \textit{info}, which can be seen in this file tree.}
\end{figure}
\bigskip 

\begin{figure*}[ht]
\centering
   \includegraphics[width=\linewidth]{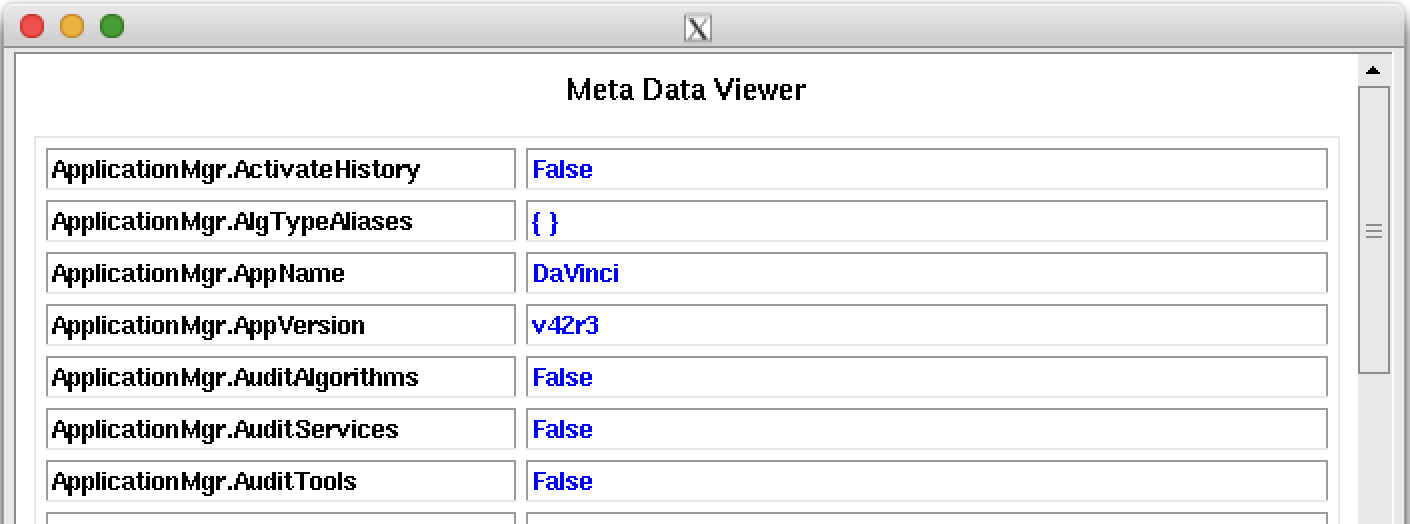}\label{fig:viewers}
   \caption{The HTML-based graphical user interface of the Metadata viewer.} 
\end{figure*}

The {\it info} file can be read in two different ways. The first one is through command-line, by simply reading in and printing the content of the dictionary. The second way is to view the content of the {\it info} file from a stand-alone provenance viewer, as shown in Figure \ref{fig:viewers}. The viewer is implemented as a pop-up window based on C++ and the \textsc{ROOT} programming language. This means that it requires the ROOT application (and an input ROOT file) to be executed. The viewer uses a table to showcase the dictionary (key-value pairs) of metadata captured in the job. It allows for a user-friendly way to see the provenance, while the command-line approach is better when there is a need to reuse the configurations or reproduce the dataset.

When it is necessary to reproduce a ROOT dataset, the original job can be recreated from the information within the dataset. This is done by extracting the metadata from the dataset and saving it in a file as a ``flat'' list of options. The list is essentially a sequence of job configurations and it will resemble the original Python application configuration file(s) created by the physics analysts. The list of configurations can be saved as a Python file, Python pickle file or Linux configuration options file. These file formats are typically used for serializing Python objects. \textsc{Gaudi} understands and processes such a file (the flat list) in the same way as the original application configuration file. Furthermore, the original application configuration file is no longer needed. Even though \DaVinci is most commonly used for physics analyses at LHCb, the Metadata service can be used for other applications within the framework that can process and produce ROOT datasets. The service was first released in \textsc{Gaudi v27r1}, meaning that it is available for use in physics analyses with  \textsc{DaVinci v40r0} onward.

\section{Conclusion}

In this paper, we introduced a new functionality of the LHCb software framework \Gaudi that captures provenance of a job and stores it directly within the output dataset. Using only the stored provenance of a dataset, the original job and the dataset can be independently reproduced. We argue that the most effective way to integrate reproducibility is through the solutions that easily incorporate into researchers' daily workflows. Our solution is implemented within an existing and widely-used scientific software. Because it does not require the analysts to conduct additional work like install third-party software or learn a new skill, it lightens the effort of enabling reproducibility in physics analyses at LHCb. In the second part of the paper, we presented our implementation and explored several different use-cases in which the provenance service would facilitate research documentation and reproducibility.

\section*{Acknowledgments}

Ana Trisovic acknowledges funding from the CERN Doctoral Student program. During her time at the University of Chicago, she was  funded in part by the Sloan Foundation.

The authors would like to thank Freddie Witherden and Patrick O'Leary, who conducted peer-review of this paper, for their insightful feedback and comments. Their comments are public and can be found at 10.22541/au.158014063.32681377 and 10.6084/m9.figshare.11536461.


\bibliographystyle{IEEEtran}
\bibliography{refs.bib}

\section*{About the authors}

\textbf{Ana Trisovic} is a Sloan postdoctoral fellow at the Institute for Quantitative Social Science, Harvard University. Her main research interests are on computational reproducibility, data provenance and open science. She completed her PhD in computer science at the University of Cambridge on the topic of data preservation and reproducibility. During that time, she worked on the LHCb experiment, CERN Open Data and CERN Analysis Preservation. Contact her at anatrisovic@g.harvard.edu and find her website at https://anatrisovic.com.

\textbf{Chris R. Jones} is a Senior Research Associate at the Cavendish Laboratory at the University of Cambridge. He completed his PhD in particle physics at Magdalene College at the University of Cambridge. He oversees the work and development of the RICH detector at the LHCb experiment. Contact him at jonesc@hep.phy.cam.ac.uk

\textbf{Ben Couturier} is a software engineer at CERN. He develops software for the LHCb experiment. He is  part of the team putting in place tools and processes to ensure software quality, with a special focus on preservation. Contact him at Ben.Couturier@cern.ch

\textbf{Marco Clemencic} is a lead software developer for the Gaudi framework at the LHCb experiment. He completed his PhD in particle physics at the University of Turin. Contact him at Marco.Clemencic@cern.ch

\end{document}